\documentstyle[ltwol,epsfig]{article}

\def\Journal#1#2#3#4{{#1} {\bf #2}, #3 (#4)}

\def\NPB{{\em Nucl. Phys.} B}
\def\PLB{{\em Phys. Lett.}  B}

\def\PRD{{\em Phys. Rev.} D}
\def\ZPC{{\em Z. Phys.} C}

\def\be{\begin{equation}}
\def\ee{\end{equation}}
\def\bea{\begin{eqnarray}}
\def\eea{\end{eqnarray}}

\begin{document}

\title{MEASUREMENTS OF $\alpha_s$ FROM HADRONIC EVENT SHAPES IN $e^+ e^-$ 
ANNIHILATION}

\author{JOHN C. THOMPSON}

\address{Rutherford Appleton Laboratory, Chilton, Didcot OX11 0QX, UK
\\E-mail: jcth@rl.ac.uk}   

\twocolumn[\maketitle\abstracts{New studies of hadronic event shape observables 
in $e^+ e^-$ collisions between 13 and 183 GeV CM energy have 
enabled the running of $\alpha_s$ to be confirmed and the validity of 
non-perturbative power-law corrections to be investigated.  A more precise value 
of $\alpha_s$(M$_Z$) with reduced theoretical errors has been reported from fitting 
18 oriented event shape distributions measured in one experiment at the $Z$.}]

\section{Running of $\alpha_s$(Q) and Power Corrections}

Most experimental measurements of $\alpha_s$ are limited in precision by 
theoretical errors arising from missing higher order terms and 
uncertainties in the hadronisation corrections which are normally derived from 
the phenomenological Monte Carlo models. 
Recently, improved perturbative QCD predictions for the event shape 
observables, C-parameter~\cite{Catani} and the Jet Broadenings~\cite{Dokshetal} 
(B$_W$, B$_T$), have become available to add to those already produced for Thrust 
(T) and Heavy Jet Mass (M$_H^2$)~\cite{DW}.
 
Leading non-perturbative (1/Q, where Q = $\sqrt s$) power corrections have also 
been calculated for these observables. 
They are based on an effective strong coupling, 
$\alpha_{0}(\mu_{I})$, at an infra-red matching scale, $\mu_{I}$ (usually set to 2 
GeV). This coupling is expected to be approximately universal but must be derived 
from experiment. Once evaluated, power corrected distributions can be compared 
directly with experiment leaving the Monte Carlo models for detector corrections 
only. Recently, a 2-loop analysis of these 1/Q power corrections has been 
performed~\cite{DLMS} for Thrust rescaling the original prediction by a so-called 
$'$Milan$'$ factor ($M$) which has been shown to apply equally to the other 
aforementioned observables~\cite{Milan}. The power corrections basically shift the 
perturbative predictions for the differential spectra of each observable linearly 
in $\alpha_{0}$ although additional logarithmic terms are thought to be required for 
the jet broadenings~\cite{Dokshetal}.     

In a contribution to this conference, ALEPH~\cite{ALEPH940} fit the power corrected 
perturbative predictions in the 3-jet regions of the 1-T, C, M$_H^2$ and B$_W$ 
distributions to the measurements at 5 values of $\sqrt s$ from 91.2 to 183 GeV. 
The perturbative predictions are ${\cal O}(\alpha_{s}^2)$ with $\ln$(R) (or R) 
matched NLL terms. The perturbative renormalisation scale parameter, 
x$_{\mu}$(=$\mu/\sqrt s$), is set to 1. They extract simultaneously values for 
$\alpha_{0}(\mu_{I})$ and $\alpha_{s}$(M$_Z$) from each observable. The fits are 
reasonably satisfactory except to the B$_W$ distributions where the results are far 
from those obtained for the other observables with either matching scheme. The bad 
quality of these fits disfavours the concept of a logarithmically enhanced 
shift~\cite{Dokshetal}. 
DELPHI~\cite{DELPHI137} and a JADE group~\cite{JADE646}, who also include data from 
higher energies, fit instead to the means of event shape distributions 
simultaneously at all $\sqrt s$ values. In this case, only the 
${\cal O}(\alpha_{s}^2)$ terms can be used 
for the perturbative contribution. OPAL~\cite{OPAL305} presented a similar analysis
applied also to the second and third moments of the 1-T and C distributions. In 
this case, the power corrections are suppressed by ($\mu_{I}$/Q)$^n$ where $n$ = 2 
or 3. Using Monte Carlo models, they show that the power corrections for the higher 
moments are small above Q = M$_Z$ and also 
a low value of x$_{\mu}$ is required to obtain consistent values of $\alpha_{s}$ 
from them. They fit simultaneously to the first three moments in the data for each 
observable at 91.2, 133, 161 and 172 GeV. When correlations are taken into account, 
the fits are not stable if x$_{\mu}$ = 1 indicating that higher order perturbative 
terms are large. Thus, x$_{\mu}$ is allowed to vary as well and is found to be 
highly correlated to $\alpha_{0}(\mu_{I})$. As expected, the results are 
insensitive to the power terms in the higher moments which are therefore set to 
zero. 

\noindent
\begin{table}
\begin{center}
\caption{Values of $\alpha_{s}$(M$_Z$) and $\alpha_{0}(\mu_{I})$ as obtained from the 
fits.
$\dagger$ Results from C and (1-T) are consistent;
$\ddagger$ values are compatible only at 30\% level.}
\label{alpha_zero}  
\begin{tabular}{|c|l|l|l|}
\hline
Expt & $\alpha_s$(M$_Z$) & $\alpha_{0}(\mu_{I})$ & x$_{\mu}$ \\
\hline 
ALEPH  & 0.1168$\pm$0.0048 & 0.451$\pm$0.065$\dagger$ & 1.0 \\
DELPHI & 0.1176$\pm$0.0057 & 0.494$\pm$0.009 (T) & 1.0 \\
       & 0.1172$\pm$0.0037 & 0.558$\pm$0.025 (C) &     \\
JADE   & 0.1188$^{+0.0044}_{-0.0034}$ & 0.325 - 0.616$\ddagger$ & 1.0 \\ 
OPAL(1)& 0.1143$\pm$0.0088 & 0.322 (T)           & 0.038 \\
       & 0.1158$\pm$0.0107 & 0.249 (C)           & 0.052 \\
OPAL(2)& 0.141             & as above            & 1.0   \\
\hline
\end{tabular}
\end{center}
\end{table}
  
Table~\ref{alpha_zero} compares the values of $\alpha_{0}(\mu_{I})$ and 
$\alpha_{s}$(M$_Z$) obtained from each experiment. The latter values are 
renormalised to M$_Z$ assuming compatibility with QCD; they are in good agreement 
except for the OPAL(2) result where x$_{\mu}$ is kept fixed. Theory systematics 
dominate and are typically evaluated from varying x$_{\mu}$ from 0.25 to 4.0 and 
$\mu_{I}$ between 1 and 3 GeV.     

\begin{figure}
\center
\epsfig{figure=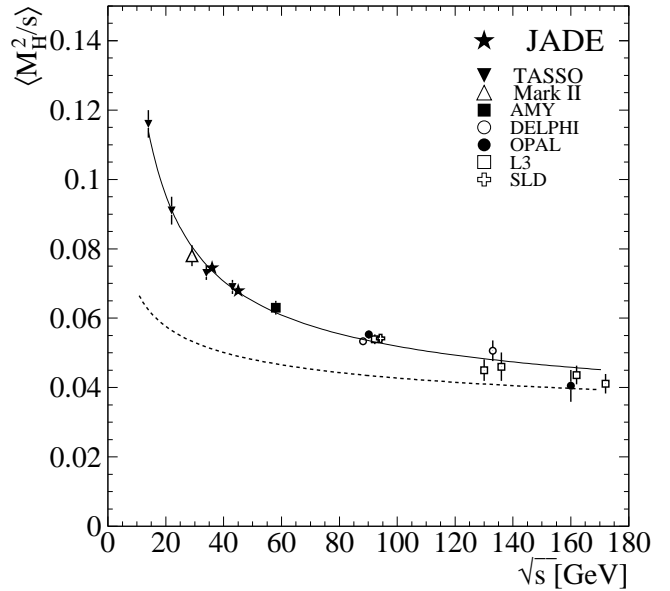,height=8.0cm}
\vskip 1cm
\epsfig{figure=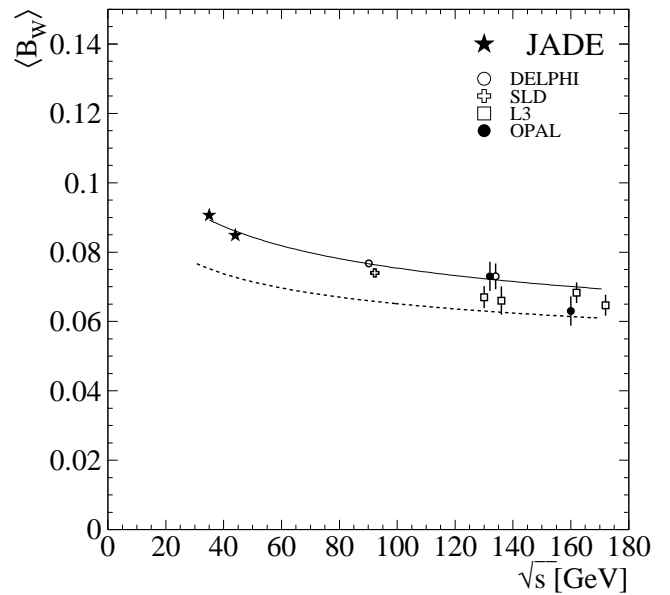,height=8.0cm}
\caption{Energy dependence of $\langle M_H^2 \rangle$ and $\langle B_W \rangle$; 
the dashed line is the perturbative prediction only.}
\label{fig:Bethke}
\end{figure}

All the LEP experiments have tested the running of $\alpha_{s}$(Q) using event shape 
observables. L3~\cite{L3-536} have fitted resummed ${\cal O}(\alpha_{s}^2)$ 
predictions of T, M$_H^2$, B$_W$ and B$_T$ to measured distributions. In this case, 
the predictions are corrected to the hadron level by Monte Carlo models. Values of 
$\alpha_{s}$(Q) are determined at 11 $\sqrt s$ energy points between 30 and 183 GeV 
using events with isolated photons to extend the energy range below the $Z$. 
The 3-loop QCD running $\alpha_{s}$ curve gives an excellent fit $\chi^2$ from 
which $\alpha_{s}$(M$_Z$) = 0.1216$\pm$0.0017$\pm$0.0058. The most precise and 
comprehensive test was submitted by a JADE group~\cite{JADE646} as described 
earlier incorporating data from other PETRA, TRISTAN, LEP and the SLD experiments. 
The 2-loop power corrected ${\cal O}(\alpha_{s}^2$) predictions for the mean values 
of 5 event shape observables as a function of $\sqrt s$ are fitted individually to 
the data over a CM energy range from 13 to 172 GeV. 
Fig.~\ref{fig:Bethke} show the best fits to the heavy jet mass and wide jet 
broadening observables. As seen by ALEPH, the fits to the jet 
broadenings have poor $\chi^2$s.
The values of $\alpha_{s}$ found are averaged with those 
from the other 3 observables, $\langle$(1-T)$\rangle, \langle$B$_T\rangle, 
\langle$C$\rangle$, to give the results quoted in Table~\ref{alpha_zero}. 
It was pointed out at the conference that the power correction terms provided for 
the jet broadenings (where fits are generally poor) are incorrect and need to be 
revised~\cite{Dokshit}. This may improve the level of consistency found.

\section{Precise determination of $\alpha_{s}$(M$_Z$) from oriented event shapes}

The DELPHI Collaboration reported a new precise determination of $\alpha_{s}$(M$_Z$) 
from a high statistics study of 1.4 million re-processed hadronic events at the $Z$
~\cite{DELPHI142}.
Eighteen event shape distributions were measured as a function of the polar angle of 
the thrust axis. The corresponding ${\cal O}(\alpha_{s}^2)$ QCD predictions were 
corrected to the hadron level using Monte Carlo models and fitted to the data in 
defined ranges individually chosen for each observable to avoid regions where these 
corrections are greater than 40\% and the acceptance less than 80\%. In addition, 
the range was adjusted until the value of $\alpha_{s}$ obtained was stable.

With the renormalisation scale fixed to x$_{\mu}$ = 1 a large scatter is observed in 
the 18 fitted values of $\alpha_{s}$ found. This arises mainly from missing higher  
order perturbative QCD contributions (resummed terms are not available for many of 
the observables used). This procedure was repeated using the 
experimentally-optimised-scale (EOS) procedure~\cite{SBethke} where x$_{\mu}$ is 
allowed to vary with $\alpha_{s}$ in a 2-parameter fit to each observable. An 
impressive reduction in the scatter is achieved although the values of x$_{\mu}$ 
required vary considerably between 0.0033 for T to 6.33 for D$_2^{Geneva}$. 
Such an improvement was not observed in a similar analysis by the SLD 
experiment~\cite{Burrows}. Although this analysis was based on only 50,000 events,
it would appear that statistics do not account for the discrepancy since 
the total errors are largely dominated by theory and hadronisation 
uncertainties in both experiments. It is likely that the choice of fit 
ranges are crucial since x$_{\mu}$ and $\alpha_{s}$ are correlated in 
the EOS procedure. Three alternative theoretically motivated schemes to determine 
values of x$_{\mu}$ for each observable were tried by DELPHI all of which reduce 
the scatter observed with the fixed x$_{\mu}$ but are not so successful as the EOS 
procedure. Table~\ref{tab:scales} shows the results using the  
 the ECH scheme~\cite{Grunberg}, the PMS scheme~\cite{Stevenson} and the BLM 
scheme~\cite{Brodsky} compared with the EOS and fixed procedures.        

\begin{table}
\begin{center}
\caption{Weighted values of $\alpha_{s}$(M$_Z$) obtained from 18 event shape 
distributions in the DELPHI experiment using various choices of renormalisation 
scales.
$\dagger$ 4 fits fail to converge and correlation with experimentally optimised 
procedure is poor.}
\label{tab:scales}  
\begin{tabular}{|c|l|l|l|}
\hline
Scale choice & $\alpha_s$(M$_Z$) & $\chi^2$/dof & x$_{\mu}$ range \\
\hline 
Optimised  & 0.1164$\pm$0.0025 & 7.3/16          & 0.0033 - 6.33 \\
fixed      & 0.1243$\pm$0.0080 & 40/15           & 1.0  \\
ECH        & 0.1148$\pm$0.0038 & 18/16           & - \\
PMS        & 0.1147$\pm$0.0040 & 21/16           & - \\
BLM        & 0.1168$\pm$0.0053 & 24/13$\dagger$  & - \\
\hline
\end{tabular}

\end{center}
\end{table}

The DELPHI submission~\cite{DELPHI142} to the conference also includes further 
studies of selected observables where resummed NLL terms to combine with the 
${\cal O}(\alpha_{s}^2)$ predictions are available. Fits are made to the data 
selecting regions of the shape distributions which are: (a) restricted to the 2-jet 
region where pure NLLA should apply and (b) to the 2+3 jet region using the full 
theory. The renormalisation scale, x$_{\mu}$, is set to 1. Acceptable fits are 
obtained to the data but in general the fits using the EOS procedure without 
resummed terms are superior over a wider range of the distributions. 

In conclusion, although apparently successful, the EOS procedure remains a 
controversial and 
somewhat unsatisfactory method of producing a precise measurement of $\alpha_{s}$. 
A better method may be to constrict the analysis to observables for which resummed 
and power-law predictions are available. A proper solution to the missing 
higher orders and other non-perturbative effects is still highly desirable.

\end{document}